\documentclass[twocolumn,showpacs,preprintnumbers,amsmath,amssymb]{revtex4}

\usepackage{graphicx}
\usepackage{dcolumn}
\usepackage{bm}
\usepackage{epsfig}

\begin{document}
\title{New approach to calculating the News}
\author{Nigel T. Bishop \& Shrirang S. Deshingkar}
\address{Department of Mathematics, Applied Mathematics \& Astronomy,
University of South Africa, P.O. Box 392, Unisa 0003, South Africa}

\begin{abstract}
We consider the problem of finding the gravitational radiation output, or
news, within the context of a numerical simulation of a spacetime by means of
the null-cone, or characteristic, approach to numerical relativity. We develop
a method for computing the news that uses an explicit coordinate
transformation to a coordinate system that satisfies the Bondi conditions.
The method has been implemented computationally. We present results of
applying the method to certain test problems, demonstrating second order
convergence of the news to the analytic value.
\end{abstract}


\maketitle

\section{Introduction}
\label{s-intro}

In many simulations in numerical relativity, an important objective is to
compute the emitted gravitational radiation. Within ADM-type approaches, one
uses computed data near the outer boundary of the simulation to estimate
the radiation at future null infinity (${\mathcal I}^+$) in the form of
an expansion of spherical harmonics. The underlying
theory was developed some time ago~\cite{regw,zeri,monc}, and the
computational implementation is discussed, for example,
in~\cite{rupar,cams,seis}. However, this paper uses the characteristic, or
null-cone, approach with Bondi-Sachs coordinates, in which ${\mathcal I}^+$ is
included in the compactified grid, and therefore the gravitational radiation,
or news, can in principle be found exactly~\cite{cce}.

Even so, the computation of the news is not a straightforward matter. The news
takes a very simple form, $N=\frac{1}{2}J_{,\ell u}$, when the null-cone
coordinate system satisfies the Bondi conditions (the meaning of these terms
and symbols will be given later). However, the Bondi conditions are conditions
at ${\mathcal I}^+$ (where $\ell=0$), and in any physical simulation the
freedoms in the null-cone coordinates are set in the interior of the
spacetime. In general, a computed metric will not satisfy the Bondi
conditions. Thus, a procedure for computing the news in a general Bondi-Sachs
coordinate system has been developed~\cite{hpn}.

A motivation for this work is that some difficulties with the news
computation of~\cite{hpn} have been reported~\cite{part}, specifically
significant gravitational radiation when the gravitational field is static. The
cause of the difficulties is not known, and it may be that they can be
resolved by technical improvements to the method and code~\cite{yosefthesis},
but as yet no such results have been reported. Our method has also not been
tested for the problem of~\cite{part}, and so it is not certain that it will
be of assistance; but because of its simplicity (see below) it would be much
easier to analyze, and thereby resolve, any problems.

However, our primary motivation is that the method presented here is much
simpler than the previous method. Conceptually, the idea is straightforward:
we construct a transformation to the ``right'' coordinate system, and
consequently there is a means to monitor the accuracy of results obtained.
The technical differences between the two methods are summarized in
Sec.~\ref{s-con}. Here we note that the present method requires two fewer
evolution equations than the previous method, and also that the most
complicated formula in the present method is Eq. (\ref{e-jl}), whereas the
formulas for the previous method were given in an appendix taking up a whole
page. In numerical work, simplicity means that there is less opportunity for
the introduction of error. Of course, the previous method has been tested
against available analytic solutions (as will be done for the method presented
here), but these solutions are rather special. Thus, whatever the 
computational performance of the two methods, it would still be necessary 
to develop the present method and use it to validate results obtained 
by the previous method.

We construct an explicit coordinate transformation
between the general Bondi-Sachs coordinate system and a Bondi-Sachs
coordinate system that satisfies the Bondi conditions (given in Eq.
(\ref{e-b})). The coordinate transformation on ${\mathcal I}^+$ is
known~\cite{hpn}, but constructing the transformation throughout the spacetime
is much more difficult. The key point of our method is that, while it is in
principle possible to find the transformation throughout the spacetime, we do
not need it. Rather, we need to evaluate $\frac{1}{2}J_{,\ell u}$ on
${\mathcal I}^+$ (where $\ell=0$), and thus we need the transformed metric off
${\mathcal I}^+$ {\em only to first order in $\ell$}. We proceed by
expressing the coordinate transformation as a Taylor series in $\ell$.
The algebraic calculations are quite lengthy and require the use of computer
algebra.

Our news calculation has been implemented computationally, and used in problems
in which an analytic result is known. Results are presented below, and it is seen
that the errors are second-order convergent to zero.

The paper starts with a summary of relevant results and notation for the
characteristic formulation of numerical relativity (Sec.~\ref{s-not}). The
coordinate transformation and computer algebra results are given in Sec.~\ref{s-coo},
and then the procedure for computing the news~-- at both analytic and computational
levels~-- is described in Sec.~\ref{s-proc}. The computational tests and results are
presented in Sec.~\ref{s-com}.
The Conclusion, Sec.~\ref{s-con}, gives a comparison between the method in this
paper and that of~\cite{hpn}; and also discusses possibilities for further work.

\section{Notation}
\label{s-not}
The formalism for the numerical evolution of Einstein's equations, in null cone 
coordinates, is well known~\cite{hpn,cce} (see 
also~\cite{ntb93,ntb90,rai83,bondi}). For the sake of completeness, 
we give a summary of those aspects of the formalism that will be used here.
We start with coordinates based upon a family of outgoing null hypersurfaces.
We let $u$ label these hypersurfaces, $x^A$ $(A=2,3)$, label
the null rays and $r$ be a surface area coordinate. In the resulting
$x^\alpha=(u,r,x^A)$ coordinates, the metric takes the Bondi-Sachs
form~\cite{bondi,sachs}
\begin{eqnarray}
   ds^2  =  -\left(e^{2\beta}(1 + {W \over r}) -r^2h_{AB}U^AU^B\right)du^2
\nonumber \\
        - 2e^{2\beta}dudr -2r^2 h_{AB}U^Bdudx^A 
        +  r^2h_{AB}dx^Adx^B,
\label{eq:bmet}
\end{eqnarray}
where $h^{AB}h_{BC}=\delta^A_C$ and
$det(h_{AB})=det(q_{AB})$, with $q_{AB}$ a unit sphere metric.
We work in stereographic coordinates $x^A=(q,p)$ for which the unit sphere
metric is
\begin{equation}
q_{AB} dx^A dx^B = \frac{4}{P^2}(dq^2+dp^2),
\end{equation}
where
\begin{equation}
        P=1+q^2+p^2.
\end{equation}
We also introduce a complex dyad $q_A$ defined by
\begin{equation}
      q^A=\frac{P}{2}(1,i), \;\;q_A=\frac{2}{P}(1,i)
\end{equation}
with $i=\sqrt{-1}$. For an arbitrary Bondi-Sachs metric,
$h_{AB}$ can then be represented by its dyad component
\begin{equation}
J=h_{AB}q^Aq^B/2,
\end{equation}
with the spherically symmetric case characterized by $J=0$. The
full nonlinear $h_{AB}$ is uniquely determined by $J$, since the
determinant condition implies that the remaining dyad component
\begin{equation}
K=h_{AB}q^A \bar q^B /2
\end{equation}
satisfies $1=K^2-J\bar J$.  We introduce the spin-weighted field
\begin{equation}
U=U^Aq_A,
\end{equation}
as well as the (complex differential) eth operators $\eth$ and $\bar \eth$
(see~\cite{eth} for full details).

The news calculation is performed in conformally compactified coordinates.
Specifically, $(u,r,x^A) \rightarrow (u,\ell,x^A)$ where $\ell = 1/r$. In
$(u,\ell,x^A)$ coordinates, the compactified metric is $d\hat{s}^2=\ell^2 ds^2$
where $\ell$ is a conformal factor with future null infinity ${\mathcal I}^+$
given by $\ell=0$. The compactified metric will be denoted by
$\hat{g}^{\alpha\beta}$, and the general compactified Bondi-Sachs metric is
\begin{eqnarray}
\hat{g}^{11}=e^{-2\beta}V_a , \;\; \hat{g}^{1A}=e^{-2\beta} U^A,
\nonumber \\
\hat{g}^{10}=e^{-2\beta}, \;\;
\hat{g}^{AB}=h^{AB}, \;\; \hat{g}^{0A}=\hat{g}^{00}=0,
\label{e-bsc}
\end{eqnarray}
where $V_a=\ell^2(1+\ell W)$. In addition to the general compactified
Bondi-Sachs metric Eq.~(\ref{e-bsc}), we will refer to the compactified
Bondi-Sachs metric satisfying the Bondi conditions, and such quantities will
be denoted with a suffix ${}_{[B]}$, with compactified metric and coordinates
$\hat{g}^{\alpha\beta}_{[B]}$ and $(u_{[B]},\ell_{[B]},x^A_{[B]})$,
respectively. On ${\mathcal I}^+$, i.e. $\ell_{[B]}=0$,
$\hat{g}^{\alpha\beta}_{[B]}$ satisfies the Bondi conditions
\begin{equation}
\hat{g}^{11}_{[B]}=0, \;\;\hat{g}^{1A}_{[B]}=0, \;\;\hat{g}^{01}_{[B]}=1, \;\;
\hat{g}^{AB}_{[B]}=q^{AB}_{[B]},
\label{e-b}
\end{equation}
where $q^{AB}_{[B]}$ is a unit sphere metric with respect to the Bondi angular
coordinates $x^A_{[B]}$.

\section{Coordinate transformation}
\label{s-coo}
We define a coordinate transformation, near ${\mathcal I^+}$, between $(u,\ell,x^A)$
and $(u_{[B]},\ell_{[B]},x^A_{[B]})$:
\begin{eqnarray}
u_{[B]}=u_{[B]0} +  A^u \ell +C^u \ell^2,\;\; \ell_{[B]} = \omega \ell + C^\ell \ell^2,
\nonumber
\\ x^A_{[B]}= x^A_{[B]0} +A^A  \ell  +C^A \ell^2, 
\end{eqnarray}
where $\omega$, $x^A_{[B]0}$, $A^A$, $u_{[B]0}$, $A^u$ are all functions of
$x^A$ and $u$ only. It will turn out that the $C^\alpha$ are irrelevant to the news
calculation, but they are
included so that the Jacobian (constructed by differentiating the above)
is manifestly correct to first order in $\ell$. We also introduce complex quantities
\begin{equation}
A=q_A A^A \mbox{ and } X = q_A x^A_{[B]0},
\end{equation}
both of which are defined to have spin weight $1$.
The general metric quantities are expanded as a Taylor series to first order
in $\ell$, with the ${}_0$ quantities being the values at $\ell=0$, i.e. on
${\mathcal I}^+$, and the ${}_\ell$ quantities being the first $\ell$ derivative
evaluated at $\ell=0$, i.e. $(\partial_\ell)_{\ell=0}$.
\begin{eqnarray}
\beta= \beta_0 +\ell \beta_{\ell}, \;\; U=U_0+\ell U_{\ell}, \;\;
J=J_0+\ell J_{\ell}, \nonumber \\
 K=K_0+\ell K_{\ell}, \;\; V_a= \ell V_{a \ell}.
\end{eqnarray}
The same first-order expansion in $\ell$ and notation will be used for the
metric quantities in the coordinate system satisfying the Bondi conditions
(e.g., $J_{[B]}=J_{[B]0}+\ell J_{[B]\ell}$). The general compactified
Bondi-Sachs metric, and the Bondi-Sachs metric satisfying the Bondi
conditions, are related by
\begin{equation}
\hat{g}^{\alpha\beta}_{[B]0}+\ell \hat{g}^{\alpha\beta}_{[B]\ell}=
\hat{g}^{\alpha\beta}_{[B]} = \omega^{-2} \frac{\partial x^\alpha_{[B]}}{\partial x^\mu}
                \frac{\partial x^\beta_{[B]}}{\partial x^\nu} \hat{g}^{\mu\nu},
\end{equation}
with the factor $\omega^{-2}$ appearing because there is also an implicit change of
compactification factor from $\ell$ to $\ell_{[B]}$.

We have used Maple to find $\hat{g}^{\alpha\beta}_{[B]}$ to first order in $\ell$, and in
doing so we have imposed the conditions (c.f. Eqs. (39) and (40) in~\cite{hpn}).
\begin{equation}
(\partial_u + U^B_0 \partial_B) x_{[B]0}^A=0
\label{eq-yevol}
\end{equation}
and
\begin{equation}
(\partial_u + U^B_0 \partial_B) u_{[B]0} = \omega e^{2\beta_0}.
\label{eq-uevol}
\end{equation}
We found that the Bondi conditions
\begin{equation}
\hat{g}^{11}_{[B]0}=0, \;\;\hat{g}^{1A}_{[B]0}=0, \;\;\hat{g}^{01}_{[B]0}=1
\end{equation}
are satisfied identically. The remaining Bondi condition is
\begin{equation}
\hat{g}^{AB}_{[B]0}=q^{AB}_{[B]}.
\label{eq-us}
\end{equation}
Taking the determinant of Eq.~(\ref{eq-us}) leads to an explicit expression
for $\omega$
\begin{equation}
\omega=  \frac{P}{P_{[B]}} \sqrt{|q_{[B],q} p_{[B],p} - q_{[B],p} p_{[B],q} |},
\label{eq-om}
\end{equation}
where $P_{[B]} = 1+ {q_{[B]}}^2 + {p_{[B]}}^2 $.
The remaining content of Eq.~(\ref{eq-us}) can be written as
\begin{equation}
J_{[B]0}= -\hat{g}^{AB}_{[B]0}q_{A[B]}q_{B[B]}/2=0,
\end{equation}
where $q_{A[B]}$ is the complex dyad appropriate to the metric $q^{AB}_{[B]}$.
The Maple calculation finds
\begin{eqnarray}
J_{[B]0}= 4P^2  {\Big[} 4J_0 X^2 \bar \zeta^2 -2 K_0 \eth X \bar \eth X + 4J_0 X 
\bar \eth X \bar \zeta   \; \; \: \; \nonumber \\
- 4 K_0 X \eth X \bar \zeta  
+ \bar J_0 (\eth X)^2
+ J_0 (\bar \eth X)^2{\Big ]} \; \; \: \; \nonumber \\
/ {[(4+P^2 X \bar X)^2 \omega^2]} =0, \; \; \: \; \; \; \: \; \; \; \: \;
\label{eq-j}
\end{eqnarray}
where $\bar \zeta = q -i p$. 
Eq.~(\ref{eq-j}) is not used to determine any of the transformation parameters, but
rather as  a check on the accuracy of the computation, as described in
Sec.~\ref{s-proc}.

In addition to satisfying the Bondi conditions, $\hat{g}^{\alpha\beta}_{[B]}$ is
also a Bondi-Sachs metric and must satisfy the conditions
\begin{equation}
\hat{g}^{00}_{[B]}=0, \;\;\hat{g}^{0A}_{[B]}=0.
\label{eq-ay}
\end{equation}
Now, $\hat{g}^{00}_{[B]0}=0$ and $\hat{g}^{00}_{[B]\ell}=0$ can be expressed as linear equations
in $A^u$ and $C^u$ respectively, but these values will not be needed. Also,
$\hat{g}^{0A}_{[B]0}=0$ and $\hat{g}^{0A}_{[B]\ell}=0$ can be expressed as linear equations
in $A^A$ and $C^A$ respectively: the expression for $C^A$ will not be needed, but
the expression for $A^A$ will be used in the news calculation and,
in complex form,
\begin{eqnarray}
A=  [J_0 \bar \eth u_{[B]} (2 X \bar \zeta + \bar \eth X) + \bar J_0 
\eth u_{[B]} \eth X  \nonumber \\ 
- K_0 ( \eth u_{[B]} (2 X \bar \zeta + \bar \eth X)+ 
\bar \eth u_{[B]} \eth X)]/(2 \omega).
\label{e-A}
\end{eqnarray}

The computer algebra calculation evaluates the inverse Jacobian at $\ell=0$, from which we
find a result that will be used in the news calculation
\begin{equation}
\frac{\partial}{\partial u_{[B]}}(u,\ell,x^A) = \frac{e^{-2\beta_0}}{\omega}(1,0,U^A_0).
\label{e-jaci}
\end{equation}

The computer algebra yields an expression for $J_{[B]\ell}$:
Defining
\begin{eqnarray}
T_1=J_0 \bar \eth X - K_0 \eth X, \nonumber \\
T_2=J_0  
({\bar \eth X})^2 + \eth X ( \bar J_0 \eth X - 2K_0 \bar \eth X ),
\end{eqnarray}
then
\begin{widetext}
\begin{eqnarray}
J_{[B]\ell} =  P^2  ( 4 (1-P_{[B]}) A T_2   +
2 X P_{[B]} (J_{\ell} (\bar \eth {X})^2 + \bar J_{\ell} (\eth X)^2 +
2 (\bar \eth A T_1
+ 2\bar \zeta X J_0  \bar \eth A   + \eth A (\bar J_0 \eth X - K_0 \bar \eth X)
- 2\bar \zeta X K_0 \eth A   \nonumber \\
-K_{\ell} \eth X  (2\bar \zeta X + \bar \eth X) )
+ 4 \bar \zeta J_{\ell} X   ( \bar \zeta X + \bar \eth X))
- 4 A X  ( -4\bar \zeta^2 J_0 X  + P_{[B]}  (e^{-2 \beta_0}
( 2\bar \zeta U_{\ell} X  + ( U_{\ell} \bar \eth X + \bar U_{\ell} \eth X ) 
+ 2 A_{,u} + \nonumber \\
A V_{a\ell} + 
( U_0 \bar \eth A + \bar U_0 \eth A ) + 2\bar \zeta A  U_0  ) 
+ 2\bar \zeta T_1 )) - P^2 X^2  \bar A  ( 4\bar \zeta X  ( \bar \zeta J_0 X  
+  T_1 ) + T_2 ) + 16 \bar \zeta A X  T_1 ) / (8X {P_{[B]}}^3  \omega^2).
\; \; \; \: \;
\label{e-jl}
\end{eqnarray}
\end{widetext}

\section{Procedure for computing the News}
\label{s-proc}

We can now describe the procedure used for calculating the news. First, we look at
the process from an analytic viewpoint, and then we give some details of the
computational implementation. As discussed in Sec.~\ref{s-intro}, we assume that
the general Bondi-Sachs metric is known throughout the spacetime.

\subsection{Analytic procedure}
\begin{enumerate}
\item Solve Eq.~(\ref{eq-yevol}) to find $x^A_{[B]0}=x^A_{[B]0}(u,x^A)$.
\item Eq.~(\ref{eq-om}) gives $\omega=\omega(u,x^A)$.
\item Use Eq.~(\ref{eq-j}) to find by how much $J_{[B]0}$ differs from zero,
and thus obtain an estimate of the accuracy of the calculation.
\item Solve Eq.~(\ref{eq-uevol}) to find $u_{[B]0}=u_{[B]0}(u,x^A)$.
\item Eq.~(\ref{eq-ay}) gives $A=A(u,x^A)$.
\item We note that
\begin{eqnarray}
\left( \frac{\partial J_{[B]}}{\partial \ell_{[B]}} \right)_{\ell_{[B]}=0}
=\frac{\partial x^\mu}{\partial \ell_{[B]}}\frac{\partial J_{[B]}}{\partial x^\mu}
\nonumber \\
=\frac{\partial \ell}{\partial \ell_{[B]}}\frac{\partial J_{[B]}}{\partial \ell}
=\frac{J_{[B]\ell}}{\omega},
\end{eqnarray}
because $J_{[B]}$ is zero on ${\mathcal I^+}$, and so derivatives in directions
other than $\ell$ vanish. Then the news $N=N(u,x^A)$ is
\begin{eqnarray}
N=\frac{1}{2} \frac{\partial^2 J_{[B]}}{\partial \ell_{[B]} \partial u_{[B]}}
=\left( \frac{\partial}{\partial u_{[B]}} \right) \frac{J_{[B]\ell}}{2 \omega}
\nonumber \\
= \left( \frac{\partial u}{\partial u_{[B]}}\frac{\partial}{\partial u}
+\frac{\partial x^A}{\partial u_{[B]}}\frac{\partial}{\partial x^A}
\right)\frac{J_{[B]\ell}}{2 \omega}.
\label{eq-n}
\end{eqnarray}
The terms $\partial u /\partial u_{[B]}$ etc. are evaluated at $\ell=0$ and were
given in Eq.~(\ref{e-jaci}) above.
\item Finally, we need to express the news as a function of Bondi coordinates, i.e.
to transform  $N(u,x^A)$ to $N(u_{[B]},x^A_{[B]})$.
\end{enumerate}

\subsection{Computational implementation}

The news module has been written to interface directly with the null gravity
code in its current form~\cite{roberto}. Thus we use a
compactified radial coordinate $x=r/(R+r)$, with $R=1$. There are $n_x$ points
in the $x$ direction in the range [0.5,~1] (corresponding to $1<r<\infty $).
We use stereographic angular coordinates, with the number of grid
points in each angular direction $n_n$.

Instead of solving equations (\ref{eq-yevol}) and (\ref{eq-uevol}),
we convert them to ordinary differential equations as in reference \cite{hpn},
and solve them by means of the second order method
\begin{eqnarray}
(x^{A(n+1)} - x^{A(n)})_i  = \Delta u \, ({U^A}_0)^{(n+\frac{1}{2})}_i ,
\label{e-xab} \\
(u_{[B]}^{(n+1)} - u_{[B]}^{(n)})_i  = \Delta u \, (\omega
e^{2\beta0})^{(n+\frac{1}{2})}_i .
\end{eqnarray}
At the initial time-slice we match the Bondi coordinates to the Bondi-Sachs
coordinates: i.e. at $u=0$, $u_{[B]}=0$ and $x^A_{[B]}=x^A$. This also  gives
$A=0$ and $J_{[B]\ell} =J_{\ell}$ on the initial surface. The solution of
Eq.~(\ref{e-xab}) leads to $x^A=x^A(u,x^A_{[B]})$, and this needs to be
inverted to give $x^A_{[B]}=x^A_{[B]}(u,x^A)$. At this stage the inversion has
not been implemented in general, because it is not needed in the test examples
considered below.

Second derivatives do not appear directly in the formulas leading to the news.
First angular (eth) derivatives are implemented using the central difference
second order method -- with second order accurate upwind/downwind at the
boundaries. The first time derivative of $A$ is needed to calculate
$J_{[B]\ell}$, and the first time derivative of $J_{[B]\ell}/\omega$
appears in the Bondi news function. For both the quantities, apart from the
first two time-steps, we use one sided- downwind second order accurate
differentiation scheme. For the initial steps we use the standard first order
scheme.

\section{Computational tests and results}
\label{s-com}

We tested our method for calculating the Bondi news 
function against two exact solutions:
linearized Robinson-Trautman, and Schwarzschild in rotating coordinates. 
We discretize this analytic solutions on the required computational
grid and determine the various metric components and use them as an input
for the news module. The news module as such
sees it as input from an equivalent numerical evolution code.

\subsection{Linearized Robinson-Trautman solution}
\label{s-rt}

The Robinson-Trautman solution represents a distorted black hole 
emitting purely outgoing radiation. The radiation decays exponentially, and 
asymptotically the solution becomes Schwarzschild.

The Robinson-Trautman metric  can be put in the form
\begin{eqnarray}
  ds^2 = -({\cal K}-{2\over r{\cal W}})du^2-2{\cal W}dudr
         -2r{\cal W}_{,A}dudx^A \nonumber \\
        +r^2q_{AB}dx^A dx^B,
\end{eqnarray}
where ${\cal K}={\cal W}^2[1-L^2(\log {\cal W})]$,
$L^2$ is the angular momentum operator (i.e., the angular part of the Laplacian
operator on the unit sphere), and ${\cal W}(u,x^A)$
has to satisfies the nonlinear differential equation
\begin{equation}
      12 \partial_u(\log {\cal W}) = {\cal W}^2 L^2 {\cal K}.
    \label{eq:rteq}
\end{equation}
We recover the Schwarzschild solution  by putting
${\cal W}=$constant in the above metric.

Linearized solutions to the Robinson-Trautman equation (\ref{eq:rteq})
are obtained by setting ${\cal W}=1+\phi$ and dropping nonlinear terms
in $\phi$,
\begin{equation}
    12 \partial_u \phi = L^2(2-L^2)\phi.  \label{eq:rt}
\end{equation}
For a spherical harmonic perturbation $\phi =\lambda(u)Y_{\ell m}$ this leads
to the exponential decay $\lambda=\lambda(0)e^{- u \,\ell(\ell+1)(\ell^2
+\ell-2)/12}$.
The (linear) perturbation can be a linear
sum of various spherical harmonics with small amplitudes. 

\begin{figure}[!]
\epsfig{file=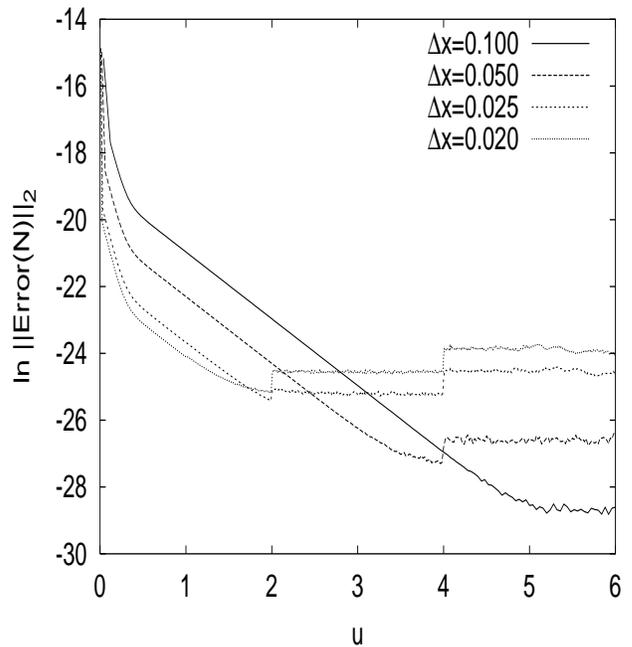,height=3.4in,width=3.5in,angle=-90}
\caption{$L_2$ norm of the error in the news as a function of time $u$ for
various discretizations for the linearized Robinson-Trautman case.}
\label{f-rtn}
\end{figure}

We consider a linearized solution of the form:
\begin{equation}
\phi = \, \Re[( \lambda_{22} e^{-2u} Y_{22} + 
	\lambda_{33} e^{-10u} Y_{33} )]
\end{equation}
with $\lambda_{22}=3\times10^{-7}$ and $\lambda_{33}=7\times10^{-7}$.
The metric components are,
$J=0$, $\beta= 0.5\phi$, $U = \eth \phi /r$ (so $U=0$ at $r=\infty$) and
$W= 1$.  Giving these quantities as inputs at the grid points at each
time level, we call the news module to calculate the Bondi 
news function. We calculate the $L_2$ norm $J_{[B]0}$ 
(which should be zero), as well as the $L_2$ norm of the error in the Bondi
news function, whose analytic value is
\begin{equation} N(u,x^A)=  \frac{1}{2}\ {\cal W}^{-1}\eth^2 {\cal W}.
\end{equation}
The computation was performed for the following grid sizes
\begin{eqnarray}
(a)& & \Delta q=\Delta p = 0.100, \Delta x = 1/60,  \Delta u = 0.04 \nonumber \\
(b)& & \Delta q=\Delta p = 0.050, \Delta x = 1/120, \Delta u = 0.02 \nonumber \\
(c)& & \Delta q=\Delta p = 0.025, \Delta x = 1/240, \Delta u = 0.01 \nonumber \\
(d)& & \Delta q=\Delta p = 0.020, \Delta x = 1/300, \Delta u = 0.008.
\label{e-grids}
\end{eqnarray}
In this case as $\zeta_{[B]} =\zeta$ we find that $||J_{[B]0}||_2$ constant 
with $u$ and
\begin{figure}[!]
\epsfig{file=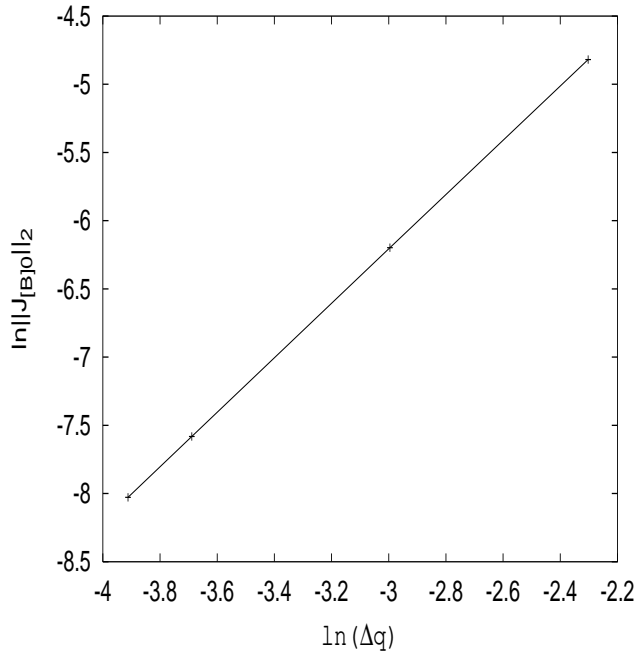,height=3.4in,width=3.5in,angle=-90}
\caption{Convergence of $||J_{[B]0}||_2$ at $u=1.6$ for the linearized 
Robinson-Trautman case.}
\label{f-rtj}
\end{figure}
 we plot $||J_{[B]0}||_2$ at $u=1.6$
for the discretizations~(\ref{e-grids}): the observed convergence rate is
$1.99$. The $L_2$ norm of the error in the news is plotted against $u$ for the
discretizations~(\ref{e-grids}) in Fig.~\ref{f-rtn}
We see convergence that, at earlier times, is approximately second order, but
at later times saturation is observed.
The convergence rate between different grids as $u$ increases is given in
Table~(\ref{Conv}).
\begin{table}[t]
\begin{tabular} {|c|c|c|c|}
\hline

convergence rate at $u $ & grids (a) \& (b) & (b) \& (c) & (c) \& (d) \\
\hline

0.4 & 1.955 & 1.992 & 1.996 \\ 
\hline

0.8 & 1.929 & 1.982 & 1.966 \\

\hline

1.2 & 1.928 & 1.966 & 1.578 \\
\hline

1.6 & 1.928 & 1.907 & 0.4746 \\

\hline

\end{tabular}

\caption{Change of convergence rate with $u$ in the $L_2$ norm of news 
function for linearized Robinson-Trautman for different grid resolutions
in equation \ref{e-grids}.}

\label{Conv}
\end{table}
Further, Fig. \ref{f-rtn}  shows that the error for
the finest grid \ref{e-grids} (d) saturates at about $e^{-24}$, 
the next finest grid \ref{e-grids}(c) 
saturates at $e^{-24.5}$, grid \ref{e-grids}(b) saturates at
$e^{-26.5}$, while  grid \ref{e-grids}(a) saturates at
$e^{-28.5}$. We observe that the saturation level scales approximately with
discretization length as $\Delta^{-3}$. These matters are discussed further in
Sec.~\ref{s-con}.

The $Y_{33}$ 
component initially dominates the total news function, which should therefore
decay approximately as $e^{-10u}$
i.e. with slope 10 on a log scale; while at later times the $Y_{22}$ component 
dominates and the news should decay as $e^{-2u}$. If the percentage of
error remains constant in the news function, the $L_2$ norm of the error 
should also show a similar trend with increasing u. We see from Fig.~\ref{f-rtj}
that this is actually the case. Until saturation effects become evident, a
little before $u=2$, the norm of the error in the news decreases as expected
at all grid resolutions.

The error in the news is significantly higher at the first two time-steps,
presumably because we are using a first order scheme since there is not enough
past data to evaluate second  order accurate time derivatives (actually, at
the second time-step one can implement a second order scheme, but its error
is larger).

\subsection{Schwarzschild in rotating coordinates}

\begin{figure}[!]
\epsfig{file=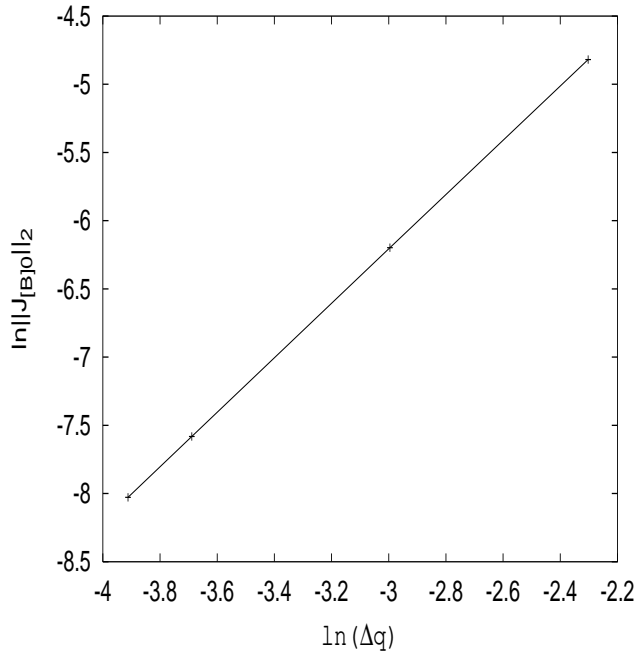,height=3.4in,width=3.5in,angle=-90}
\caption{Convergence of $||J_{[B]0}||_2$ at $u=1.6$ for the rotating 
Schwarzschild case.}
\label{f-sj}
\end{figure}

\begin{figure}[!]
\epsfig{file=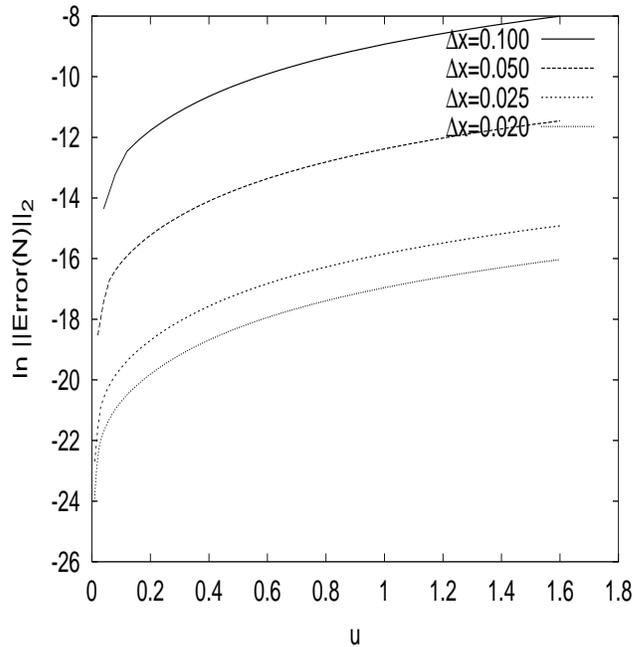,height=3.4in,width=3.5in,angle=-90}
\caption{$L_2$ norm of the error in the news as a function of time $u$ for
various discretizations for the rotating Schwarzschild case.}
\label{f-sn}
\end{figure}

By the transformation $\varphi \rightarrow \tilde \varphi + \kappa u$ of the
azimuthal coordinate, the Schwarzschild line element in null
coordinates can be written as
\begin{eqnarray}
   ds^2=-(1 - {2 m \over r}-\kappa^2 r^2 \sin^2\theta  )du^2 -2 dudr
\nonumber \\
       +2\kappa r^2 \sin^2 \theta dud\varphi
       +r^2q_{AB}dx^Adx^B.    \label{eq:smet}
\end{eqnarray}
 In this case we set
$\kappa = 4$ 
This coordinate change gives a nontrivial value for $U$ at ${\cal
I}^+$ and thus  it becomes a useful test to check that the
numerically calculated Bondi news function remains zero. 
In this case the the values of the various metric components
are $\beta = 0$, $J=0$, $U = 2I\kappa \zeta / (1+\zeta \bar\zeta) $, $W=1$.
Giving these as inputs we calculate $||J_{[B]0}||_2$ and the
news function $||N||_2$, both of which have zero as the analytic value.
We again found that $||J_{[B]0}||_2$ hardly varies with $u$, and in 
Fig.~\ref{f-sj}
 we plot $||J_{[B]0}||_2$ at $u=1.6$
for the discretizations~(\ref{e-grids}): the observed convergence rate is
again $1.99$. The $L_2$ norm of the news is plotted against $u$ for the
discretizations~(\ref{e-grids}) in Fig.~\ref{f-sn}.
We find convergence to zero at  $u=1.6$ at a rate $4.97$ averaged 
over the grids~(\ref{e-grids}) (with only minor variations according to the 
specific grid used).  The higher than expected rate is presumably due to
some cancellation effect in the truncation errors.
It is interesting that the magnitude of the error in the news builds up quite
rapidly with time. Since the error remains convergent to zero, this must mean
that the truncation error constants are growing with time.

\section{Conclusions}
\label{s-con}

We have presented a method for calculating the news, based upon an explicit
coordinate transformation from general Bondi-Sachs coordinates to a
Bondi-Sachs coordinate system that satisfies the Bondi conditions. The method
has been implemented computationally and validated against known analytic
solutions. There are similarities and differences between the method and that
of~\cite{hpn} \begin{itemize}
\item Both methods require $J=0$ on ${\mathcal I}^+$ at $u=0$.
\item Both methods solve evolution equations for $x^A_{[B]0}$ (Eq.~(\ref{eq-yevol}))
and $u_{[B]0}$ (Eq.~(\ref{eq-uevol})).
\item The method of~\cite{hpn} solves evolution equations for $\delta$
(Eq.~(\cite{hpn}-36)) and $\omega$ (Given after Eq. (\cite{hpn}-31)). In our
method, $\delta$ is not required, and $\omega$ is found explicitly
(Eq.~(\ref{eq-om})).
\item The formulas used here are much simpler than those of~\cite{hpn}.
\item {\bf There is a means to monitor the accuracy of results obtained}: 
we simply find $J_{[B]0}$ and compare the value to zero.
\end{itemize}

The saturation and reduction in convergence rate reported in Sec.~\ref{s-rt},
is typically an indication that round-off error becomes significant as the
magnitude of the signal decays. The saturation level scales approximately as
$\Delta^{-3}$, which corresponds to the finite difference representation of a
third derivative. This is consistent with the fact that here the news 
calculation implicitly involves third derivatives: $x^A_{[B]0}$ and 
$u_{[B]0}$ are differentiated to obtain $A$ (Eq.~\ref{e-A})); then $A$ is 
differentiated to obtain $J_{[B]\ell}$ (Eq.~(\ref{e-jl})); and finally 
$J_{[B]\ell}$ is differentiated to obtain the news $N$ (Eq.~(\ref{eq-n})). 
The observed scaling of saturation as $\Delta^{-3}$ is for the specific
case of the Robinson-Trautman solution, which has the special property
$x^A_{[B]0} = x^A_0$; thus, in a general case the effect may be 
more severe. For future applications, it is important to be aware that 
round-off error may be a problem.

It will be interesting to apply our news calculation to the problem considered 
in~\cite{part} (for which the method of~\cite{hpn} was unsuccessful), but that is
deferred to further work. Also deferred to further work is the possibility of
extending the coordinate transformation to higher order in $\ell$, and so being
able to extract the Bondi mass and angular momentum aspects~-- the difficulty will
be that the computer algebra is likely to lead to very complicated formulas.

\section*{Acknowledgements}
We benefited from the hospitality of the Max-Planck-Institut f\"ur 
GravitationsPhysik, Albert-Einstein-Institut and  the Center for 
Gravitational Wave Physics at Pennsylvania State University 
(funded by the NSF cooperative agreement PHY-0114375). The work 
was supported by the National Research Foundation, South Africa under Grant 
number 2053724. We thank Sascha Husa and Jeffrey Winicour for discussions.

\bibliography{v3}

\end{document}